# Coherent combining in an Yb doped double core fiber laser

**Johan Boullet, David Sabourdy, Agnès Desfarges-Berthelemot, Vincent Kermène,
Dominique Pagnoux and Philippe Roy**
IRCOM, 123 avenue Albert Thomas, 87060 Limoges, France
**Bernard Dussardier, Wilfried Blanc**
*LPMC/FOA, Université de Nice Sophia-Antipolis, Parc Valrose, 06108 Nice, France*

**Abstract**: Coherent combining is demonstrated in a clad pumped Yb doped double core fiber laser. A slope efficiency of more than 70 % is achieved with 96 % of the total output power on the fundamental mode of one of the two cores. This high combining efficiency is obtained when both cores are coupled via a biconical fused taper in a Michelson interferometer configuration.

## 1. Introduction

Thanks to their compactness, good heat dissipation, high conversion efficiency, and high beam quality, fiber lasers are very attractive sources especially for free space communications, range finding, or manufacturing applications. However, to ensure fundamental mode laser emission, the transverse dimension of the fiber core is limited to a few micrometers, involving high spatio temporal confinement of the signal wave. The consequence is that, due to the non linear effects, the spatial and spectral qualities of the laser beam may be dramatically damaged.
To increase the non linear effect threshold, several solutions have already been experimented: Large Mode Area (LMA) fiber lasers [1-3], taper performing modal filtering [4], beam clean up by wave mixing in non linear materials [5]. An other attractive way consists in designing clad pumped multicore fiber lasers [6,7]. Because of a large overlap between doped cores and pump radiation, high pump absorption is achieved with these fiber lasers. Moreover, the distribution of the power density in several cores leads to a significant reduction of non linear effects. Up to now, in this kind of composite structures, the emission occurs on a supermode with a large M² compared to the fundamental Gaussian beam.

Recently, we have developed a very efficient combining method of single mode fiber lasers using a multi-arm resonator [8]. Several groups have studied this configuration for different operating regimes and for a large number of arms [8-12]. In this paper, we propose an evolution of this method to realize an original fiber laser architecture, based on a clad pumped fiber with two Yb doped cores. The targeted goal is to combine the radiations from the two cores into the fundamental gaussian mode at the fiber laser output.

## 2. Power combining principle

The principle of the coherent combining method is based on laser self organization properties ensuring emission of modes of lowest loss, and on the use of an interferometric resonator configuration [9]. To coherently combine two independent fiber lasers, we spliced two fiber amplifiers, each one ended by a high reflectivity mirror, to the input arms of a 2X2 3dB fiber coupler for building a Michelson interferometer. One output end of the coupler is perpendicularly cleaved for providing the light feedback (4% Fresnel reflection), whereas the opposite one is angle cleaved to prevent any feedback. The coherent coupling of the fields propagating into the two amplifier arms is performed by the coupler: only the frequencies leading to constructive interferences on the shared arm can oscillate. In other words, the reflectivity transfer function of such a passive interferometer can be written: $R(\omega)=\sin^2(\omega.\Delta L/c)$ where $\omega$ is the pulsation of the optical wave, c is the light velocity, and $\Delta L$ the difference in effective length between the two arms of the interferometer. Assuming an equal gain on each arm, the interferometer acts as a spectral filter which period is $\Delta\nu=c/(2.\Delta L)$ and the gain profile provided by the set up is modulated by $R(\omega)$, with a maximum modulation depth. Owing to the self organisation property of the laser, only the longitudinal modes with minimum loss in the shared arm can oscillate in the cavity. We have shown [9] that this kind of laser based on an interferometric configuration, allowed to obtain a 3dB gain in power with regard to a single arm fiber laser. This high combining efficiency is obtained provided that the two beams interfering onto the coupler have the same polarization state. This is the reason why a polarization controller (PC) is set in one arm of the interferometer. Moreover, since an optical path difference $\Delta L$ between the two arms exists with $\Delta L > \dfrac{\lambda_0^2}{2\Delta B}$ ($\lambda_0$: central wavelength, $\Delta B$: laser bandwidth), the laser undergoes no power fluctuations despite the interferometric configuration.

This combining method still operates with two amplifying guides embedded into the same component. Fig. 1 shows a double core fiber for which coupling is locally ensured by a few millimetre long fused biconical coupler, located near one end of the fiber. At this end, a differential feedback is applied to the two cores, whereas a high reflectivity mirror reflects light towards both cores at the opposite end.

Efficient power combining may be achieved by controlling the polarization state within the laser, thanks to the use of a two core birefringent fiber. In these conditions, the laser oscillates on the two polarization eigenmodes, and the compactness of the structure is preserved. For a laser bandwidth $\Delta B$, the propagation constant difference $\Delta\beta$ between the modes of the two cores must verify the following relation: $\dfrac{\Delta\beta}{\beta_0} > \dfrac{\lambda_0^2}{2L_0\Delta B}$ for providing a large enough optical path difference between the two arms of the interferometer. $\beta_0$ is the average propagation constant of the fundamental mode in both cores. $L_0$ represents the average optical length of the resonator arms. In our experiment (Yb amplifying media) where $\Delta B \sim 20$ nm, $\lambda_0 \sim 1075$ nm, $L_0 \sim 7$ m, the minimal propagation constant difference is $\Delta\beta \sim 4.10^{-6}.\beta_0$. The very low required discrepancy $\Delta\beta$ is simply obtained thanks to the unavoidable slight differences between the optogeometrical characteristics of the two cores induced along the manufacturing process. Then, it is not necessary to draw a fiber with specific different cores.

## 3. Experiments and results:

We have designed and manufactured a fiber with two parallel Yb doped cores which axes are spaced by a distance d=20µm sufficient to prevent any coupling between the two cores. The arrangement of the two cores in the fiber is shown in the inset of fig.1. Both cores are similar and elliptical, their major and minor axes being respectively a=5µm and b=3.7µm; their numerical aperture is 0.13. The fiber was coated with a low refractive index polymer providing an inner cladding numerical aperture of about 0.4 with a diameter of 70 μm.

The experimental setup is shown on the figure 1. A 10 mm long coupler was performed near one end of the two core fiber to achieve a power exchange of about 50% at 1075 nm. The fiber end near the coupler was cleaved with an angle of about 15°, in order to prevent any Fresnel reflection. The laser was closed by a 5% output mirror $M_2$ that can be oriented to perform a selective feedback towards only one core of the fiber (core 1 on Fig. 1). At the opposite end, the 5.3m long Yb doped double core fiber was end pumped through a dichroïc mirror $M_1$ set in contact with the end of the fiber (Tmax@980nm; Rmax@1064nm) for achieving light feedback towards both cores.

With identical light feedback at the laser output, two beams of equal power were emitted (Fig. 2a). When mirror $M_2$ performed a differential feedback, a large part of the beam power was transferred from the core with no feedback to the other one (Fig. 2b and 2c). At the same time, we have verified that two beams of equal power level were emitted at the opposite end of the two core fiber through $M_1$. The performances of the two core fiber laser in terms of energy are reported on figure 3: 96% of the total output power is emitted from the core where the feedback is performed with a slope efficiency higher than 70% (core 1 in the configuration of the fig. 1). Only 4% is measured at the output of the other core (core 2). These results show that efficient coherent combining is achieved at the laser output. As expected, we have observed stable output power emission despite the interferometric behaviour of the device because the modulation period Δλ is lower than the laser bandwidth.

As in previously published experiments [8,13], the interferometric architecture of the two core fiber laser leads to a spectral modulation of the combined output beam so that the modulation depth is 1 (Fig. 4). We observed such a spectral behaviour with the two core fiber configuration. Recorded spectra of Fig. 4 exhibit a modulation period Δλ close to 0.51 nm corresponding to $\Delta L/L_0 \sim 0.22 \cdot 10^{-3}$ and $\Delta\beta/\beta_0 \sim 4 \cdot 10^{-4}$. Fig. 4a shows that the spectra of the radiations simultaneously emitted by cores 1 and 2 exhibit identical modulations when the power combining is achieved. By switching the feedback from one core to the other, the emitted spectrum is shifted by a half period (Fig. 4b) due to the Michelson geometry of the resonator.

When power combining was performed with two distinct fiber lasers [9], a random shift of the spectral modulation inside the laser bandwidth was observed due to fluctuations of the optical path difference ΔL between the two arms of the active interferometer, induced by environmental perturbations. In the two core fiber laser, both guides suffer nearly the same external perturbations, as they are embedded in the same cladding and the position of the spectral modulation remains stable.

In a previously published work [13], we have shown that the use of polarization dependent coupler for realizing the coherent combining of light makes any active control of the polarization unnecessary. The beam splitter used to coherently combine two Nd:YAG bulk lasers allowed only linearly polarized modes to oscillate in a 3-mirror cavity. In the two core fiber laser, the two elliptical cores are birefringent with a beat length close to 20 mm. Moreover, we have experimentally verified that the coupling ratio of the coupler was significantly polarization dependent. Then, as in [13], only eigenmodes with lowest loss can oscillate in the cavity and efficient power combining is achieved without any additional control of the polarization.

**Conclusion:**

We have experimentally demonstrated coherent combining in an Yb doped double core fiber laser. Up to 96% of the total output power is combined into the fundamental mode of one of the two cores, with a slope efficiency higher than 70%. This high efficiency could be still improved by orienting the main axis of the two cores in the same direction [14]. This efficient power combining is confirmed by the expected modulated spectrum of the emitted radiation. In such a configuration, the power is distributed over in a large area that significantly increases the nonlinear threshold, as well as it decreases the pump absorption length.
.
In order to extend this combining method for obtaining higher power emission, fibers with larger cores and lower numerical aperture should be used. Moreover, as the method was already applied to efficiently couple 8 distinct lasers [15], it should be extended to a fiber with a large number of active cores. High efficiency coherent combining should be obtained in this case provided that the two following conditions are fulfilled: on the one hand, the distance between neighbouring cores should be sufficient for forbidding any light exchange upstream and downstream the coupler and on the other hand a minimum difference should exist between the propagation constants in each core. In these conditions, high power single mode emission should be achieved with such a multicore fiber.

Figure captions

1. Experimental set-up of the two doped core fiber laser and transverse geometry of the fiber. YbDDCF: ytterbium doped double core fiber.

2. Output beam profile of the two core fiber laser: a) without output mirror, b) with an optical feedback performed on the core 2, c) with an optical feedback performed on the core 1.

3. Output power versus absorbed pump power.

4. Output spectrum from: a) core 1 and 2 when coherent combining is performed on core 1, b) core 1 when coherent combining is performed on core 1, and core 2 when coherent combining is performed on core 2.

**Figure**

**Figure 1:**

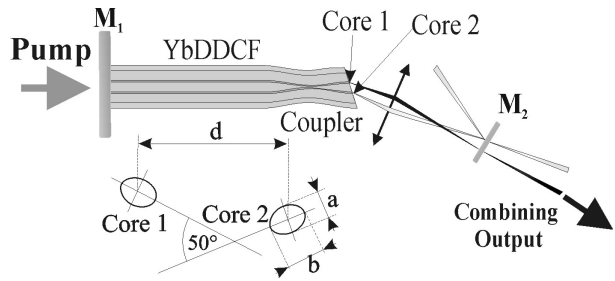

**Figure 2.**

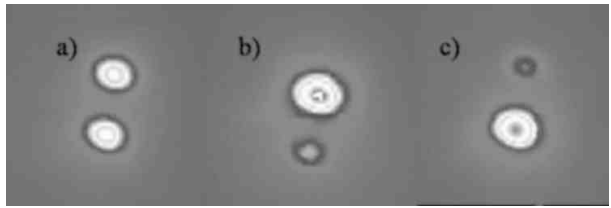

**Figure 3.**

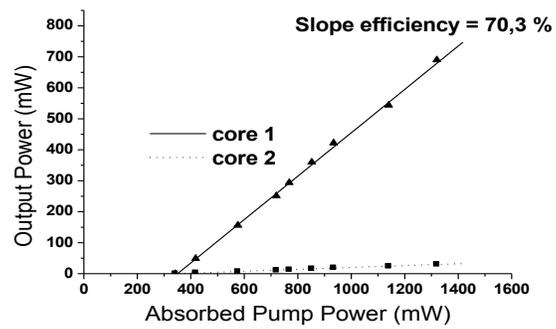

**Figure 4.**

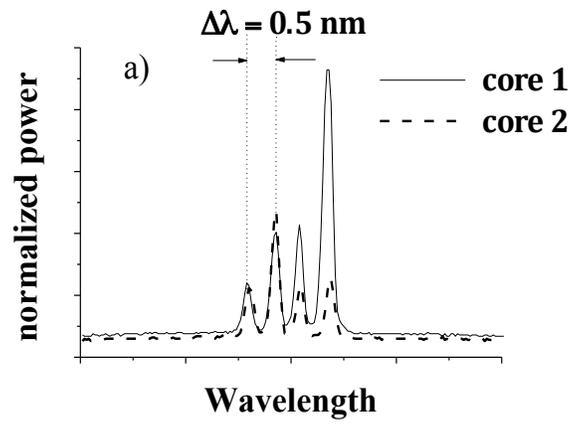

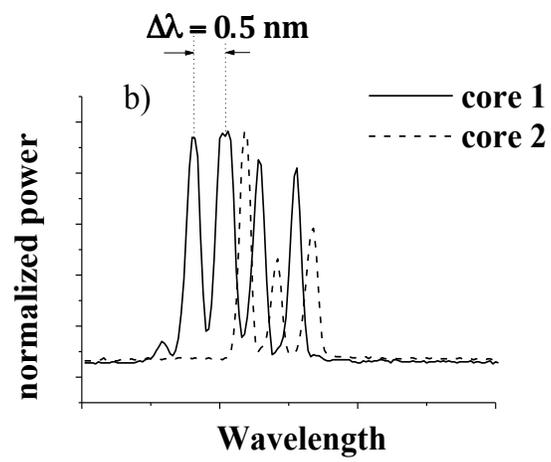